\documentclass[final,1p,times]{elsarticle} 
\usepackage{graphicx}
\usepackage{amssymb} 
\usepackage{amsthm} 
\usepackage{lineno}

\usepackage{wasysym}

\newcommand{\pT}			{\ensuremath{p_{\mathrm{T}}}}
\newcommand{\mT}			{\ensuremath{m_{\mathrm{T}}}}
\newcommand{\nq}			{\ensuremath{n_{\mathrm{q}}}}
\newcommand {\sqrtSnn}		{\ensuremath{\sqrt{s_{\mathrm{NN}}}}}

\newcommand {\tev}			{\mbox{${\rm TeV}$}}
\newcommand {\gev}			{\mbox{${\rm GeV}$}}
\newcommand {\mev}		{\mbox{${\rm MeV}$}}
\newcommand {\mmom}		{\mbox{\rm MeV$\kern-0.15em /\kern-0.12em c$}}
\newcommand {\gmom}		{\mbox{\rm GeV$\kern-0.15em /\kern-0.12em c$}}
\newcommand {\mmass}		{\mbox{\rm MeV$\kern-0.15em /\kern-0.12em c^2$}}
\newcommand {\gmass}		{\mbox{\rm GeV$\kern-0.15em /\kern-0.12em c^2$}}
\newcommand {\fmoverc}		{\mbox{\rm fm$\kern-0.15em /\kern-0.12em c$}}
\newcommand{\Tch}			{\ensuremath{T_{\mathrm{ch}}}}
\newcommand{\Tc}			{\ensuremath{T_{\mathrm{c}}}}
\newcommand{\muB}		{\ensuremath{\mu_{\mathrm{B}}}}
\newcommand{\Tkin}			{\ensuremath{T_{\mathrm{kin}}}}
\newcommand{\meanBetaT}	{\ensuremath{\langle \beta_{\mathrm{T}}\rangle}}

\journal{Nuclear Physics A} 

\begin{document}

\begin{frontmatter} 

% Your Title - please insert
\title{Global variables and correlations: Summary of the results presented at the Quark Matter 2012 conference}

%% Multiple authors
\author[auth1]{Boris Hippolyte}
\address[auth1]{Institut Pluridisciplinaire Hubert Curien, D\'{e}partement de Recherches Subatomiques,  23 rue du Loess, F-67037 Strasbourg et Universit\'{e} de Strasbourg, France}
\author[auth2]{Dirk H.\ Rischke}
\address[auth2]{Institut f\"ur Theoretische Physik, Goethe University, Max-von-Laue-Str.\ 1, D-60438 Frankfurt am Main}

\begin{abstract} 
  In these proceedings, we highlight recent developments from both theory and experiment related
  to the global description of matter produced in ultra-relativistic heavy-ion collisions as presented
  during the Quark Matter 2012 conference.
\end{abstract} 

\end{frontmatter} % do not change

%% linenumbers are useful for reviewing process
%% \linenumbers

%%-----------------------------------------------------------------------------
\section{Introduction}
  Describing the matter produced in ultra-relativistic heavy-ion collisions with a limited set of
  global variables is a tantalising task.
  In fact, the challenge is not only a description of the matter created under such extreme
  conditions but also an understanding of the details of its evolution.
  It is fair to say that stunning progress has occurred in the last couple of years, as it is
  reflected in the latest results presented during this edition of the Quark Matter conference.
  A significant fraction of the discussions focused on the shape of the initial energy density of
  the collision in terms of fluctuations.
  Such a picture is now constrained by extremely precise measurements of the azimuthal
  anisotropy of emitted particles as discussed in Sec.~\ref{sec:fluid_dynamics}.
  A fairly detailed ``standard model'' for the dynamical evolution of a heavy-ion
  collision emerges and is explained in Sec.~\ref{sec:modelling}.
  Recently, lattice QCD calculations have made tremendous progress in eliminating systematic
  uncertainties.
  As explained in Sec.\ \ref{sec:lattice}, various groups now agree on the value of the critical
  temperature for the chiral transition within systematic and statistical uncertainties, and on the
  value of the interaction measure at high temperature to within 25\%.
  Section~\ref{sec:chemical_thermal} is dedicated to recent experimental estimates of chemical
  and thermal freeze-out variables: statistical thermal analyses are performed on hadron abundances
  and comparisons are made for nuclei and hyper-nuclei as well.
  Baryon-to-meson ratios in the intermediate $\pT$ region are studied at RHIC and LHC in order
  to investigate hadronisation mechanisms involving parton recombination.
  Blast-wave fits to hadron $\pT$-spectra are used to extract thermal freeze-out conditions as a
  function of collision centrality and beam energies.
  We conclude our review of global observables and fluctuations with an outlook towards
  further key questions that need to be addressed in the near future.
%%-----------------------------------------------------------------------------
  \section{Constraints on fluid dynamics: azimuthal anisotropy}
  \label{sec:fluid_dynamics}
\begin{figure}[t]
  \begin{center}
  \includegraphics[width=.33\linewidth]{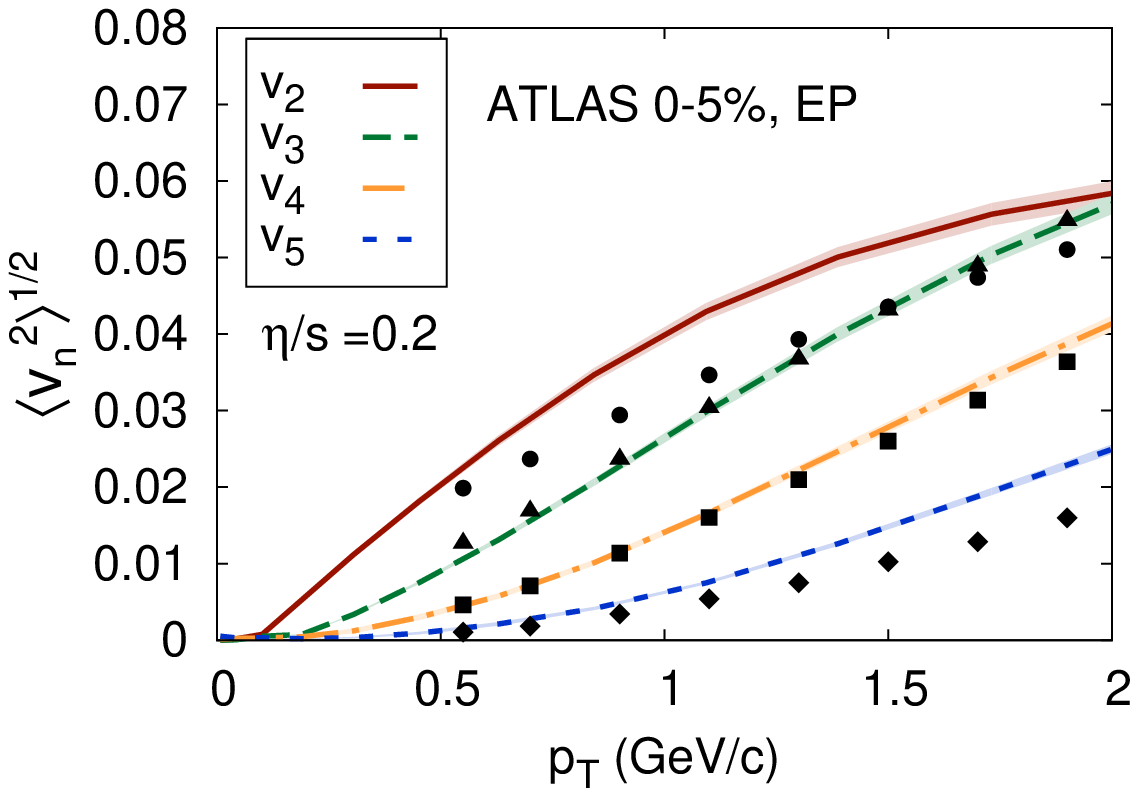}
  \includegraphics[width=.33\linewidth]{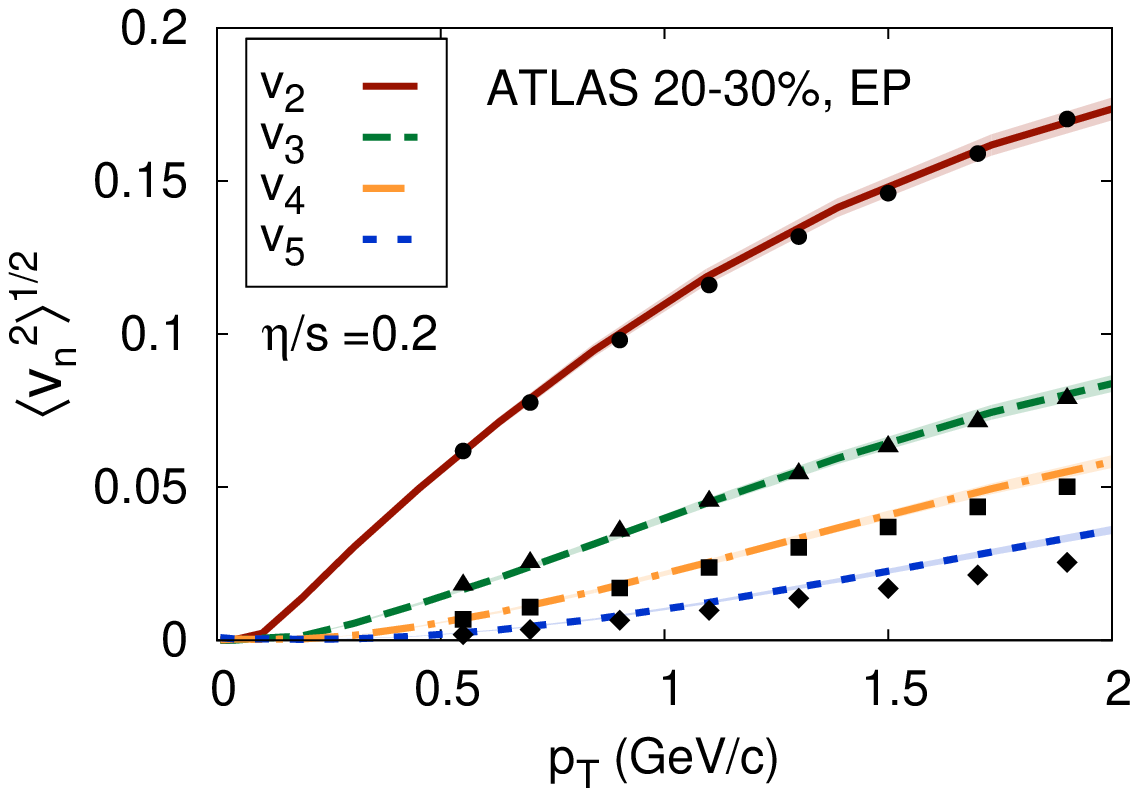}
  \includegraphics[width=.32\linewidth]{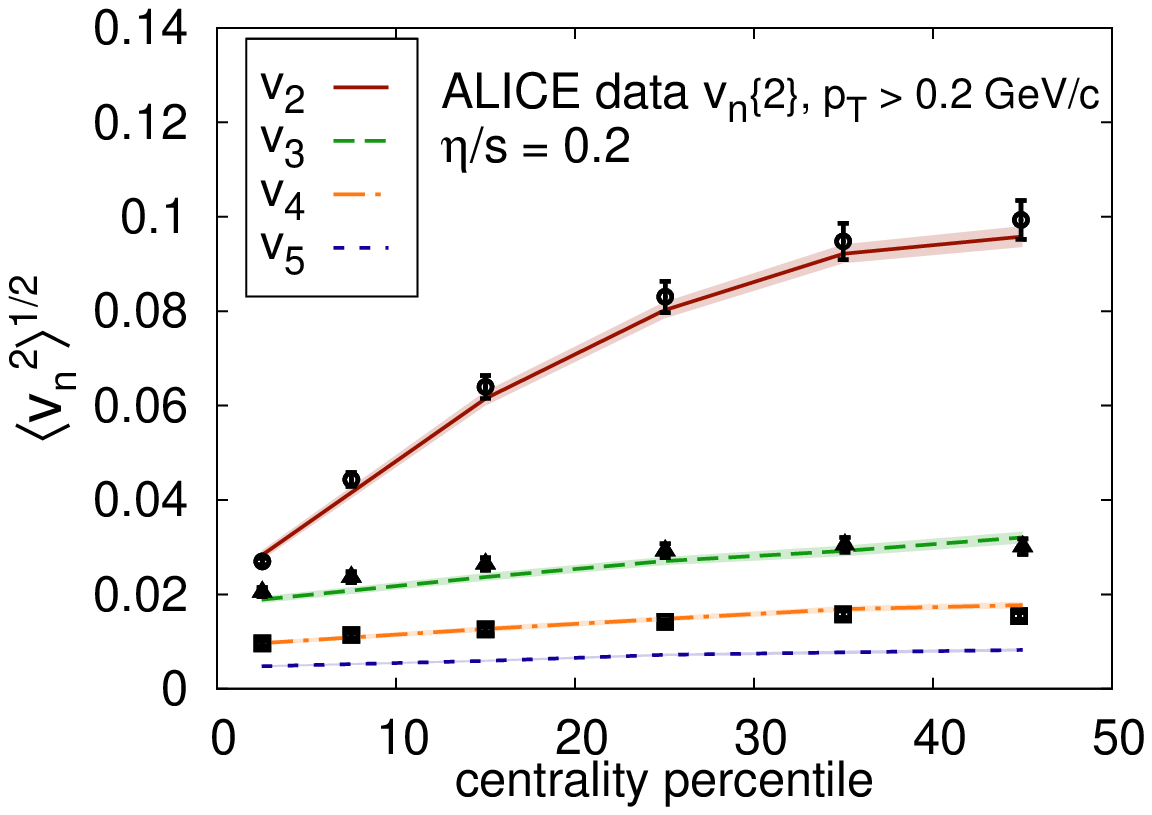}
  \end{center}
  \caption{Harmonic flow coefficients for charged particles: left and middle panel as a function of $\pT$
  for two different centrality classes~\cite{ATLAS:2012at}, right panel as a function of centrality
  percentile~\cite{ALICE:2011ab}. Calculations from Refs.~\cite{Gale:2012in,Gale:2012rq}.}
  \label{fig:1}
  \end{figure}

  The fluid-dynamical description for the bulk dynamics of heavy-ion collisions received a
  tremendous boost in interest with the measurement of elliptic flow at RHIC, showing that fluid
  dynamics could quantitatively describe the collective flow of matter and coining the paradigm of
  the Quark-Gluon Plasma (QGP) being the ``most perfect liquid'' ever created.
  Later, fluctuation measurements were perceived as a handle on both initial state as well as transport
  properties of matter produced (first as a strongly interacting QGP -- sQGP -- and then as a hadron gas).
  Currently, we are at the crossroads where state-of-the-art modelling of initial conditions meets
  extremely precise experimental measurements of fluctuations.
  This is spectacularly illustrated by the presentation of B.~Schenke with the ``real (conference)
  time\footnote{The limited statistics at the time of the presentation was later improved, 
  see~Ref.~\cite{Gale:2012in}.}"
  matching of the ATLAS event-by-event (EbyE) flow fluctuations for the anisotropic flow coefficients
  $v_{n=2,3,4}$~\cite{Jia:2012v} by the Impact-Parameter Saturated (IP-Sat) Glasma modelling  of
  the initial conditions~\cite{Schenke:2012wb}, followed by 3+1--dimensional EbyE relativistic viscous
  fluid-dynamical evolution with MUSIC~\cite{Ryu:2012at}.
  One of the striking features is that both $\pT$ and centrality differential $v_{n=2,3,4,5}$ distributions
  from ATLAS~\cite{Jia:2012v} and ALICE~\cite{ALICE:2011ab}, respectively, are correctly reproduced
  with a single value of the shear viscosity over entropy density ratio $\eta/s=0.2$ (see Fig.~\ref{fig:1}).
  %%--------------------
  %%--------------------
  Such a value is consistent with the quantification of the systematical uncertainties performed by M.\ Luzum
  and for which the ``conservative" range of $0.07 \le \eta/s \le 0.43$ is obtained~\cite{Luzum:2012wu}.
  Nevertheless, it is a major achievement that IP-Sat~Glasma initial conditions manage to reconcile $v_{2}$
  and  $v_{n\ge3}$ contrarily to MC-Glauber or MC-KLN initial energy density distributions for LHC energies.
  Moreover, with ultra-central collisions ($2\permil$) recorded and selected by CMS~\cite{Tuo:2012qm}
  as well as with the studies of correlations between measured event-plane angles, one can expect new
  challenges for models and other insights into the fluctuations of the initial energy density.
  The main conclusions from the extraction of $v_{n}$ with the Beam Energy Scan (BES) data at RHIC are
  similar but bring further constraints: when comparing to models, a low $\eta/s$ is always favoured (i.e., close
  to $0.16$) but the $\pT$-differential $v_{3}$ measured by STAR seems to increase progressively up to
  $\sim30\%$ with increasing beam energy $\sqrtSnn=11.5\rightarrow200~\gev$~\cite{Pandit:2012mq}.
  With the addition of particle identification ($\pi^{\pm}$, $\mathrm{K}^{\pm}$, $\mathrm{p}$, and
  $\mathrm{\overline{p}}$) up to $\pT \simeq 4~\gmom$ at $\sqrtSnn=200~\gev$, PHENIX reports an ordering
  for $v_{n=3,4}$ with the same pattern than what is seen for $v_{2}$ and no beam energy dependence for
  $\pT$-differential $v_{n=3,4}$ for each particle species in a 20--60$\%$ centrality interval~\cite{Gu:2012br}.
  Although the analyses require more statistics, a significant effort is made by the experimental collaborations
  at RHIC and LHC to perform $v_{2}$ estimates with higher cumulants in order to isolate non-flow and
  fluctuation contributions.
  In parallel of the well-known fact that the $\pT$-integrated $v_{2}$ increases by $\sim30\%$ from top
  RHIC to LHC energies~\cite{Aamodt:2010pa,Chatrchyan:2012t,Bold:2012qm}, it is underlined that the 
  $\pT$-differential $v_{2}$ extracted with the 4-particle cumulant method stays close in the range
  $1\le \pT \le 3~\gmom$ from lower BES RHIC  to LHC energies~\cite{Shi:2012ba}.

  Concerning the LHC studies, it is also reported that
  (i) for all collision centralities, the measurements of integrated $v_{1}$ and $v_{3}$ do not change if one uses
  $4$- or $6$-particle cumulant methods (similarly to what was noticed for $v_{2}$);
  (ii) although $v_{3}$ shows also an increase with $\pT$, and then saturates around $\pT \simeq 3~\gmom$, its
  centrality dependence is smaller than the one of $v_{2}$ (to be expected if initial density fluctuations are indeed
  at the origin of  $v_{3}$)~\cite{Bilandzic:2012an}.
  It is important to mention here the excellent agreement between different experiments when comparing $v_2$
  obtained with the $4$-particle cumulant method~\cite{Bold:2012qm}.
  
   %%--------------------
  \begin{figure}[t]
  \begin{center}
  \includegraphics[width=.9\linewidth]{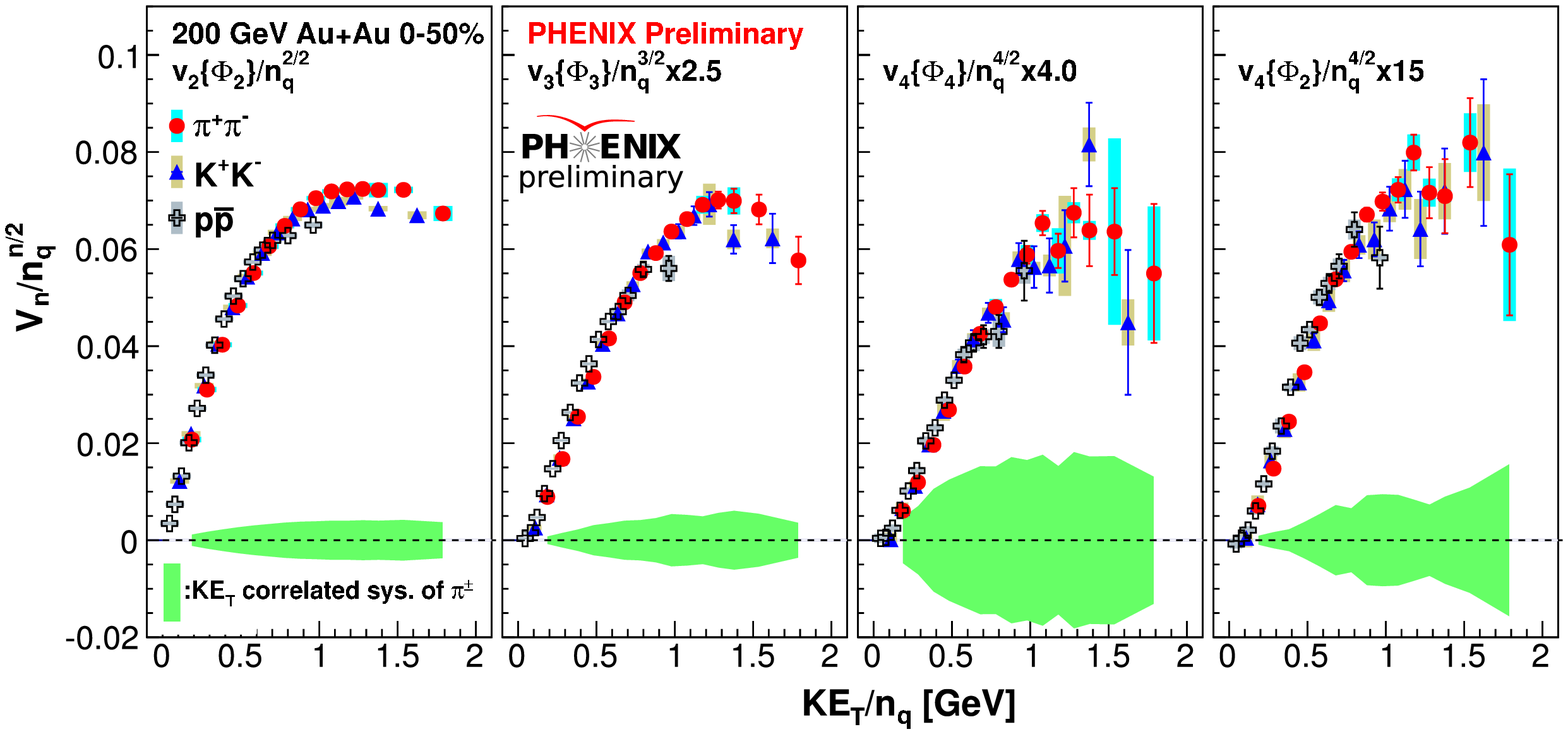}
  \end{center}
  \caption{$v_{n}/\nq^{n/2}$ vs.\ $KE_{\mathrm{T}}/\nq$ with $n=$~2--4 (from left to right) for $\pi^{\pm}$,
  $\mathrm{K}^{\pm}$, $\mathrm{p}$+$\mathrm{\overline{p}}$, in central (0--50$\%$) Au--Au collisions at
  $\sqrtSnn=200~\gev$~\cite{Gu:2012br}.}
  \label{fig:ncq_scaling}
  \end{figure}
  %%--------------------  
  Several possibilities were investigated for finding a universal scaling for produced hadrons as a function
  of their number of constituent quarks ($\nq$) : (i) as shown in Fig.~\ref{fig:ncq_scaling}, PHENIX presented
  $v_{n}/\nq^{n/2}$ vs.\ $KE_{\mathrm{T}}/\nq$ for $\pi^{\pm}$, $\mathrm{K}^{\pm}$, 
  $\mathrm{p}$+$\mathrm{\overline{p}}$
  which seems to hold for several harmonics (i.e., $n=$~2--4 for Au--Au collisions at $\sqrtSnn=200~\gev$)
  and several collision energies~\cite{Gu:2012br}; (ii) STAR reported that $v_{2}/\nq$ vs.\ $(\mT-m_{0})/\nq$
  works for most identified particles and up to top RHIC energies except an intriguing 2-$\sigma$ deviation
  for the $\phi$ meson at the lowest energies of $\sqrtSnn=7.7~\gev$ and $11.5~\gev$~\cite{Shi:2012ba}; (iii) no
  scaling seems to hold up to LHC energies~\cite{Bilandzic:2012an}.
%%-----------------------------------------------------------------------------
  \section{State-of-the-art modelling of heavy-ion collision dynamics}
  \label{sec:modelling}
\begin{figure}[t]
   \begin{center}
   \begin{minipage}{9cm}
   \vspace*{-2.8cm}
   \includegraphics[width=3cm,angle=-90]{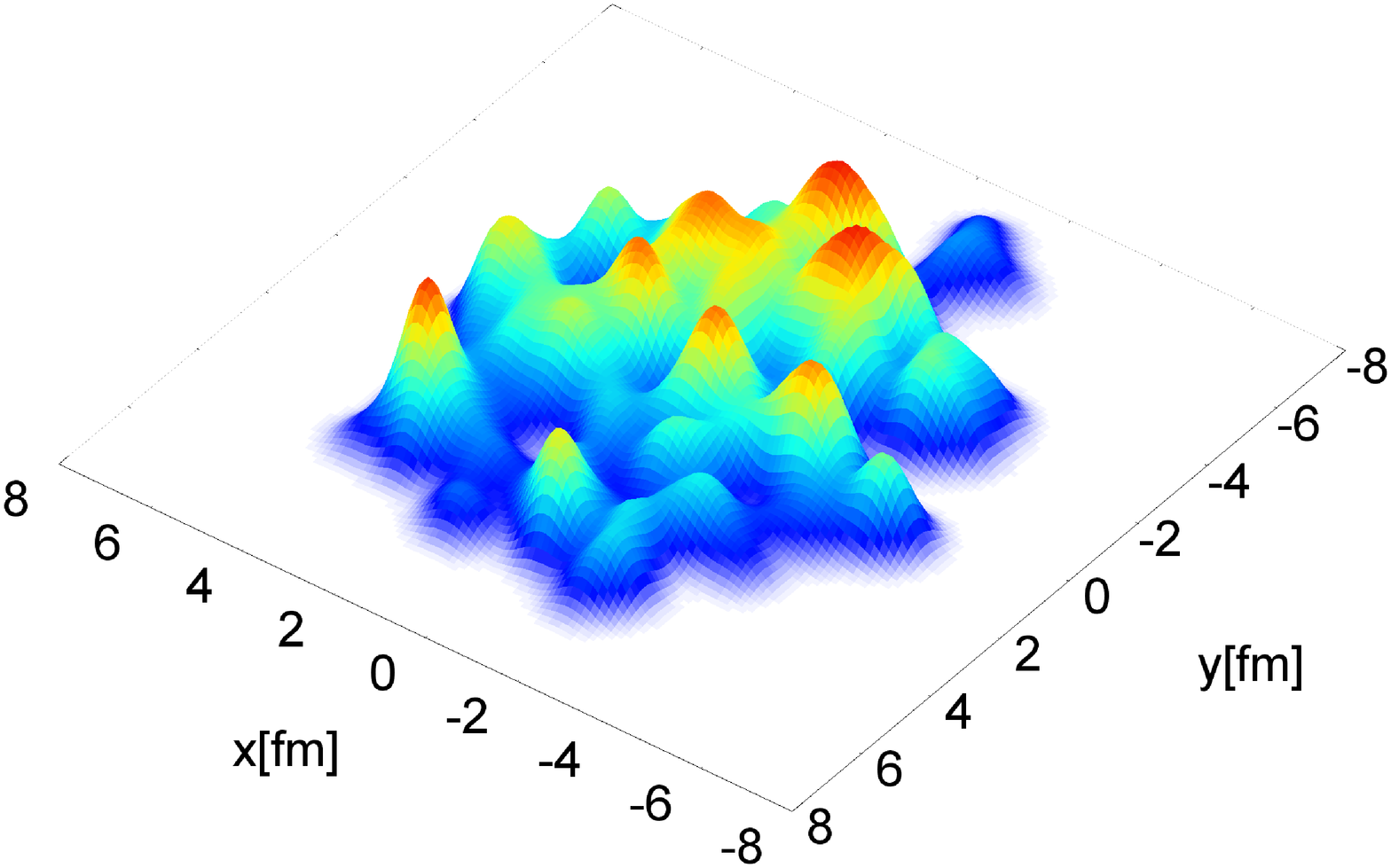}
   \includegraphics[width=3cm,angle=-90]{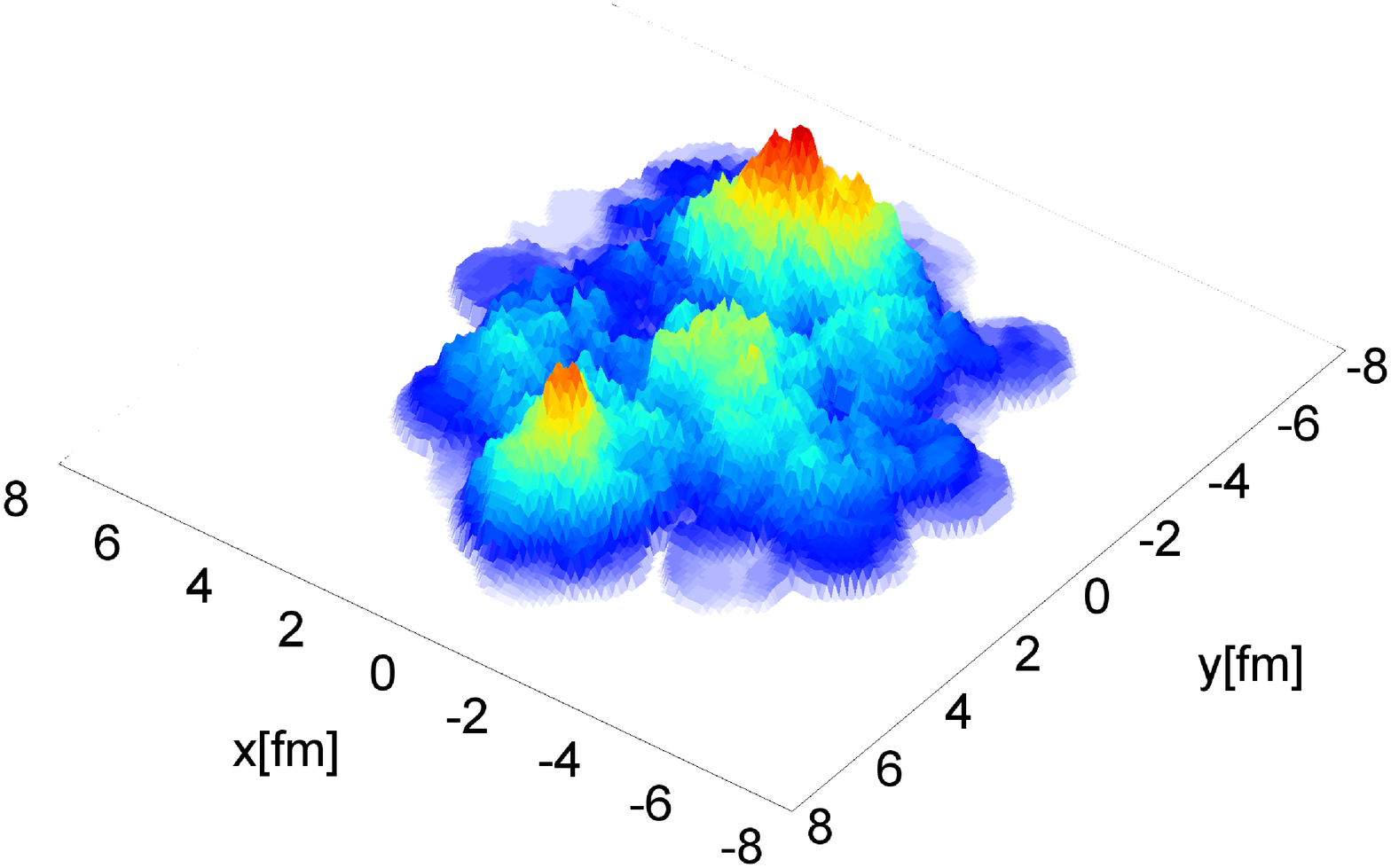}
   \end{minipage}
   \includegraphics[width=4cm]{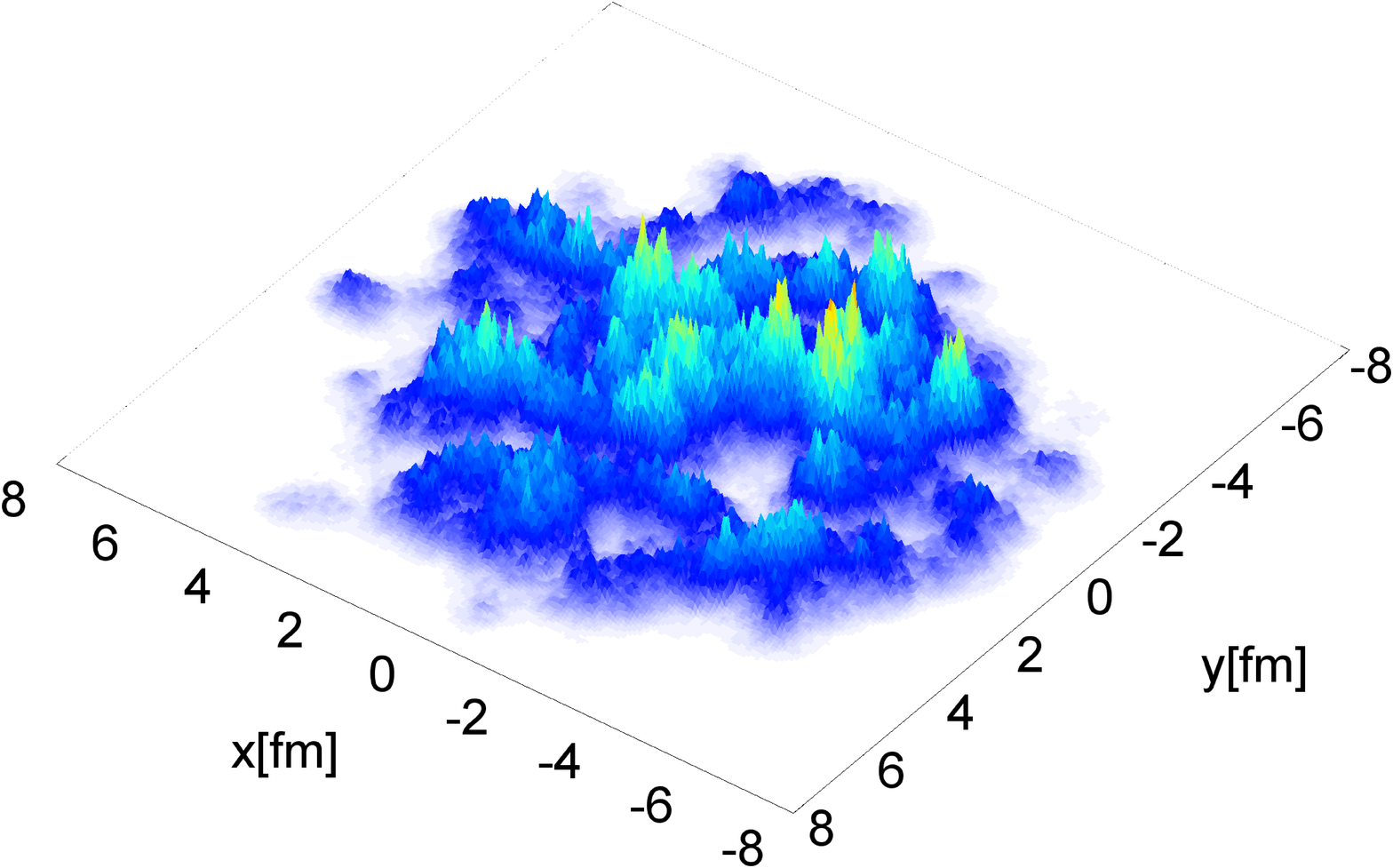}
   \end{center}
   \caption{Initial conditions in the Glauber, the KLN, and the IP-Sat model, from Refs.~\cite{Gale:2012in,Gale:2012rq}.}
   \label{fig:inicond}
   \end{figure}

  The ``standard model'' for the dynamical evolution of hot and dense strongly interacting matter created in
  ultra-relativistic heavy-ion collisions is the following: (i) individual parton-parton collisions copiously create
  gluons and quarks.
  The initial points of production are computed within various approaches.
  Some possibilities are the standard Glauber model for nucleon-nucleon collisions, the KLN model~\cite{Kharzeev:2004if},
  or the IP-Glasma model~\cite{Gale:2012in,Schenke:2012wb} which combines the Impact-Parameter Saturation
  (IP-Sat) model for the initial nuclear wave functions \cite{IPSat} with the  classical Yang-Mills dynamics of the
  produced color fields (``Glasma'')~\cite{Glasma}.
  In a Monte-Carlo implementation of these models, initial conditions fluctuate from event to event.
  For an illustration, we show the transverse energy density profile at a time $\tau = 0.2~\fmoverc$
  after the collision in Fig.~\ref{fig:inicond}.
  The IP-Sat model shows fluctuations on the smallest scales, followed by the KLN model.
  The Glauber model has the smoothest initial conditions, i.e., the smallest fluctuations.
   %%--------------------
   %%--------------------
   How these initial conditions evolve further and on which time scale thermalization is achieved is presently not
   clear.
   Calculations within $\phi^4$ theory \cite{Dusling:2012ig} show that actually thermalization (i.e., establishing an equation
   of state) is achieved prior to isotropization (i.e., equality of transverse and longitudinal pressure), cf.~left panel
   of Fig.~\ref{fig:phi4_intmeas}.
   This may necessitate the use of anisotropic fluid dynamics \cite{Florkowski:2012as}.

   Commonly, one assumes a rather rapid approach to local thermodynamical equilibrium, so that fluid dynamics
   is applicable. The initial conditions for fluid dynamics are then specified on a hypersurface in space-time
   (commonly a constant proper time surface $\tau = \tau_0 = const.$).
   From then on, the evolution of the system is determined by the conservation laws for net-charge and
   energy-momentum, 
   \begin{equation}
   \partial_\mu N^\mu = 0\;, \;\;\; \partial_\mu T^{\mu \nu} = 0\;.
   \end{equation}
   For a unique solution, causal and stable formulations of dissipative fluid dynamics require in addition
   dynamical equations for the dissipative components of $N^\mu$ and $T^{\mu \nu}$.
   For a systematic derivation of these equations from kinetic theory, see Ref.~\cite{Denicol:2012cn}.
   If the microscopic interaction rates drop below the macroscopic expansion rate of the fluid, a fluid cell will
   not be able to sustain (approximate) local thermodynamical equilibrium any longer; the cell ``freezes out''.
   This freeze-out process is commonly performed along a hypersurface of constant temperature~\cite{Cooper:1974qi}.
   How to compute the single-inclusive particle spectra in the presence of viscous terms was also reported at
   this conference~\cite{Denicol:2012qm}.
   The subsequent evolution of the system is either performed via a microscopic ``afterburner'' which takes
   into account elastic (and possibly inelastic) collisions between individual particles before they hit the detector
   (see, e.g.~Refs.~\cite{Song:2010aq,Schenke:2010nt}), or one simply assumes complete cessation of microscopic interactions.
   Then, one simply computes the single-inclusive particle spectra along the freeze-out hypersurface in order
   to compare with experimental data. Despite the apparent crudity of the underlying assumptions, the latter
   procedure is remarkably successful in reproducing the measured harmonic flow coefficients for charged
   particles, cf.\ Fig.~\ref{fig:1} where they are shown for a calculation within relativistic dissipative fluid
   dynamics~\cite{Gale:2012in,Gale:2012rq} with IP-Glasma initial conditions and a constant shear
   viscosity-to-entropy density ratio of $\eta/s = 0.2$.
   One observes nearly perfect agreement for all flow coefficients for all centrality classes, except the most central
   one where the elliptic flow coefficient $v_2$ somewhat exceeds the experimental values.
%%-----------------------------------------------------------------------------
  \section{New lattice QCD results}
  \label{sec:lattice}
\begin{figure}[t]
  \begin{center}
  \includegraphics[width=0.47\linewidth]{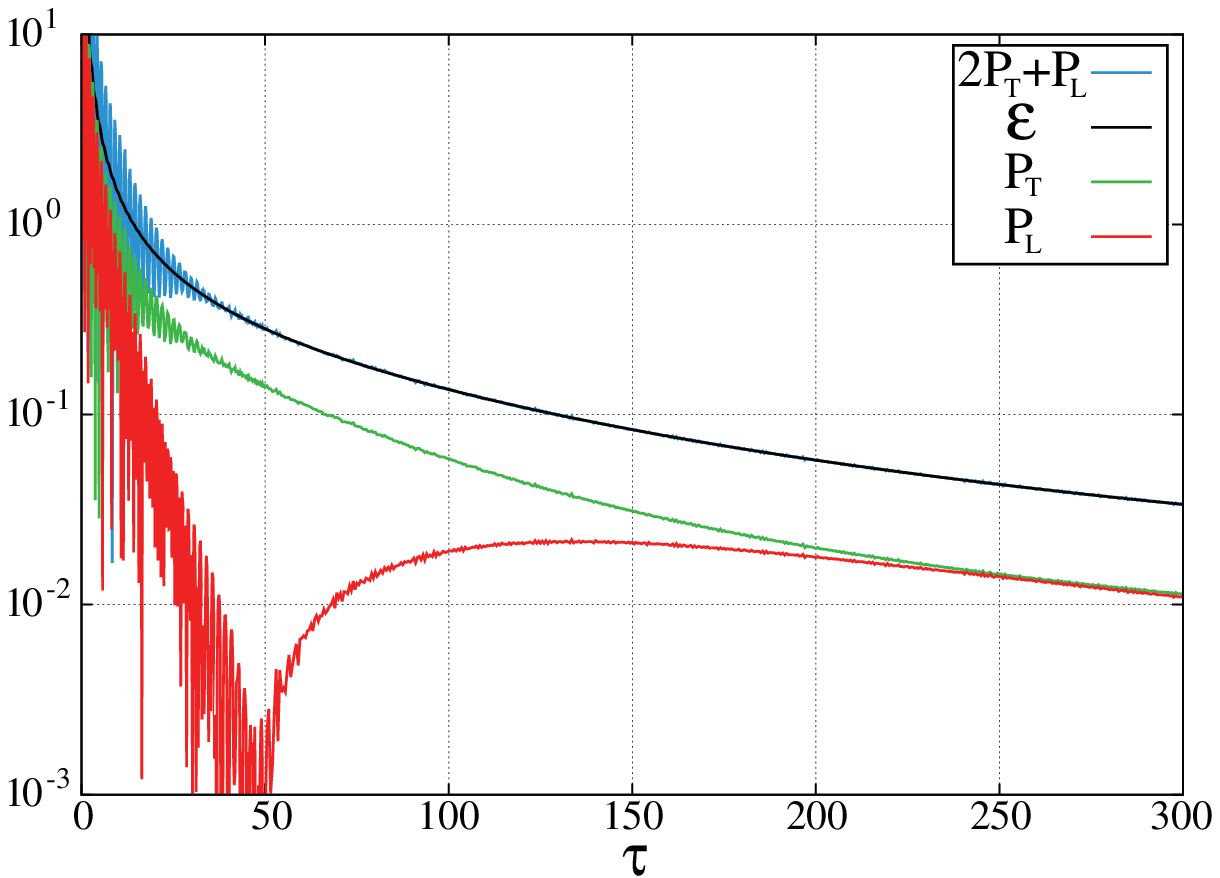}
  \includegraphics[width=0.50\linewidth]{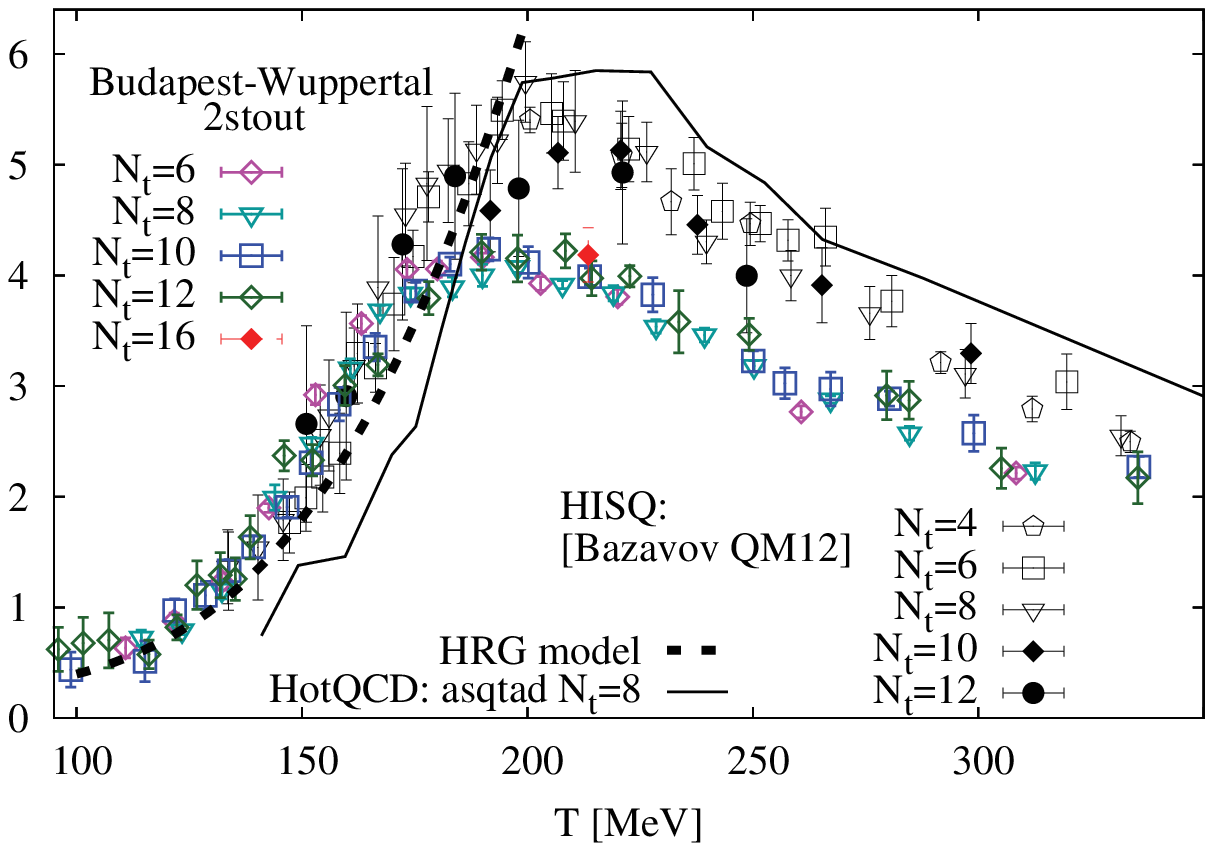}
  \end{center}
  \caption{Left: evolution of various components of the energy-momentum tensor within $\phi^4$ theory, from
  Ref.~\cite{Dusling:2012ig}. When $2P_\mathrm{T} +P_\mathrm{L}$ (blue) approaches the energy density (black), thermalization is achieved,
  while isotropization happens when the transverse (green) and longitudinal (red) pressure become degenerate.
  Right: the interaction measure $(\epsilon- 3p)/T^4$ from various lQCD calculations \cite{Borsanyi:2012rr}.}
  \label{fig:phi4_intmeas}
  \end{figure}

  As outlined in the former section, there remains a large uncertainty in the choice of initial conditions for
  fluid-dynamical calculations.
  Thus, there is little hope in being able to perform a reliable determination of the equation of state (and probably
  also of the transport coefficients) from experimental flow data.
  In this situation, lattice QCD (lQCD) remains the most important, and moreover quantitatively reliable, tool to provide
  microscopic input for fluid-dynamical calculations.
  There has been considerable progress in the past year: lQCD calculations from the HotQCD and Wuppertal-Budapest 
  collaborations now agree on the value of (pseudo-)critical temperature for chiral symmetry restoration:
  $\Tc = 154 \pm 8~({\rm stat.}) \pm 1~({\rm sys.})~\mev$~\cite{Bazavov:2011nk} vs.\ $\Tc = 155 \pm 3~({\rm stat.}) \pm 3~({\rm sys.})~\mev$~\cite{Borsanyi:2010bp}.
  Regarding the equation of state, one of the most important (systematic) uncertainties that remain to be resolved
  resides in the interaction measure.
  As can be seen from Fig.~\ref{fig:phi4_intmeas} (right), while the values of the HotQCD collaboration agree fairly
  well with those of the Wuppertal-Budapest collaboration below $\Tc$, they are systematically larger by about 25\%
  above $\Tc$.

  Another important input in dissipative fluid-dynamical calculations are the transport coefficients.
  One would hope that, in the future, lQCD calculations of these quantities reach a similar precision as those for the
  equation of state.
  Until then, one has to rely on more phenomenological approaches to extract values for $\eta/s$ and the other
  coefficients, see e.g.~\cite{Kovtun:2011np,Hidaka:2009ma}.
  %%--------------------
  %%--------------------
  
\begin{figure}[htbp]
  \begin{center}
  \includegraphics[width=5cm]{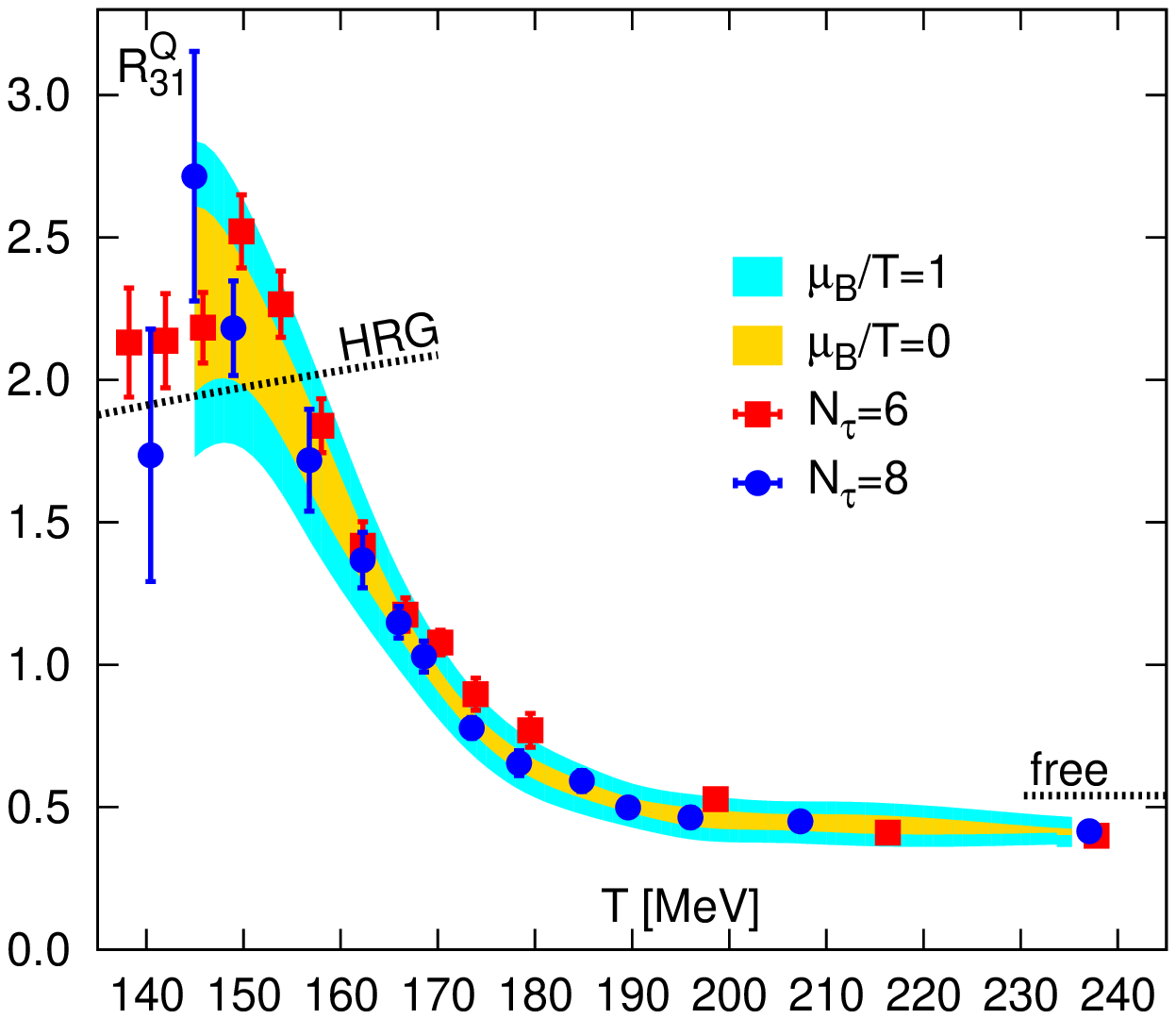}
  \includegraphics[width=5cm]{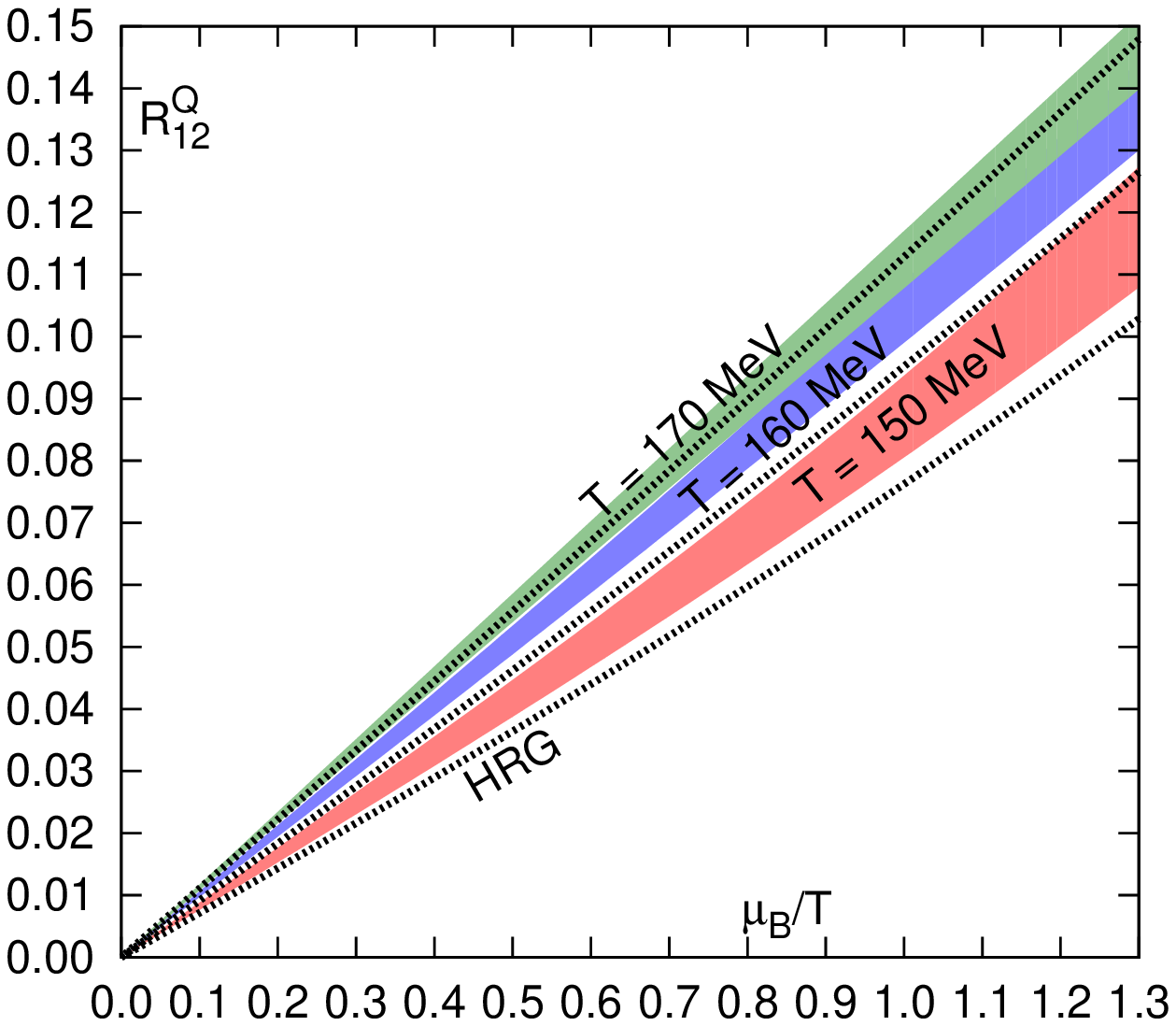}
  \end{center}
  \caption{$R_{31}^Q$ as function of $T$ (left) and $R_{12}^Q$ as
  function of $\mu_B/T$ (right), from Ref.\ \cite{Bazavov:2012vg}.}
  \label{fig:RijQ}
  \end{figure}

  Correlations between, resp.~fluctuations of, various quantities  are an important tool to learn about critical behavior
  near phase transitions.
  This conference has seen a plethora of new data for these quantities.
  What is most striking is that lQCD calculations are now in the position to make firm predictions that can be directly
  compared to experimental data.
  Figure~\ref{fig:RijQ} shows $R_{31}^Q = \langle \delta N_Q^3 \rangle/ \langle N_Q \rangle$ (left panel) and
  $R_{12} ^Q = \langle N_Q \rangle/ \langle \delta N_Q^2 \rangle$ (right panel), where $N_Q$ is the net charge
  and $\delta N_Q = N_Q - \langle N_Q \rangle$ its deviation from its average~\cite{Bazavov:2012vg}.
  $R_{31}^Q$ is fairly insensitive to the value of the baryochemical potential $\mu_B$, so it can serve as a
  thermometer to determine the (chemical) freeze-out temperature.
  Once the temperature has been extracted, $R_{12}^Q$, being fairly insensitive to temperature, can be used to
  extract the value of $\mu_\mathrm{B}/T$ at (chemical) freeze-out.
  %%--------------------
  %%--------------------
%%-----------------------------------------------------------------------------
\section{Further constraints on chemical and thermal freeze-out}
  \label{sec:chemical_thermal}
\begin{figure}[t]%[thbp]
  \begin{center}
  \includegraphics[width=.31\linewidth]{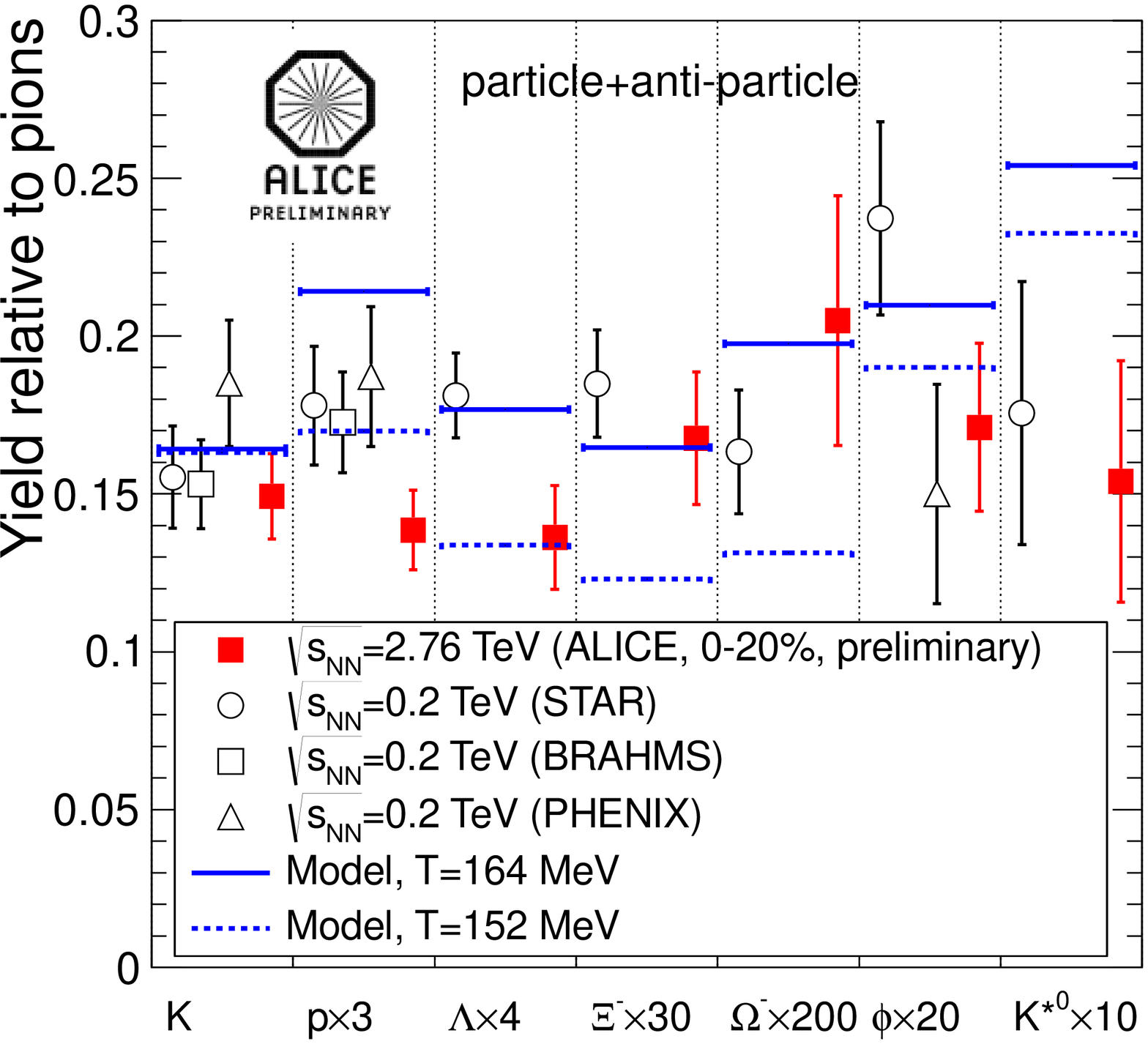}
  \includegraphics[width=.35\linewidth]{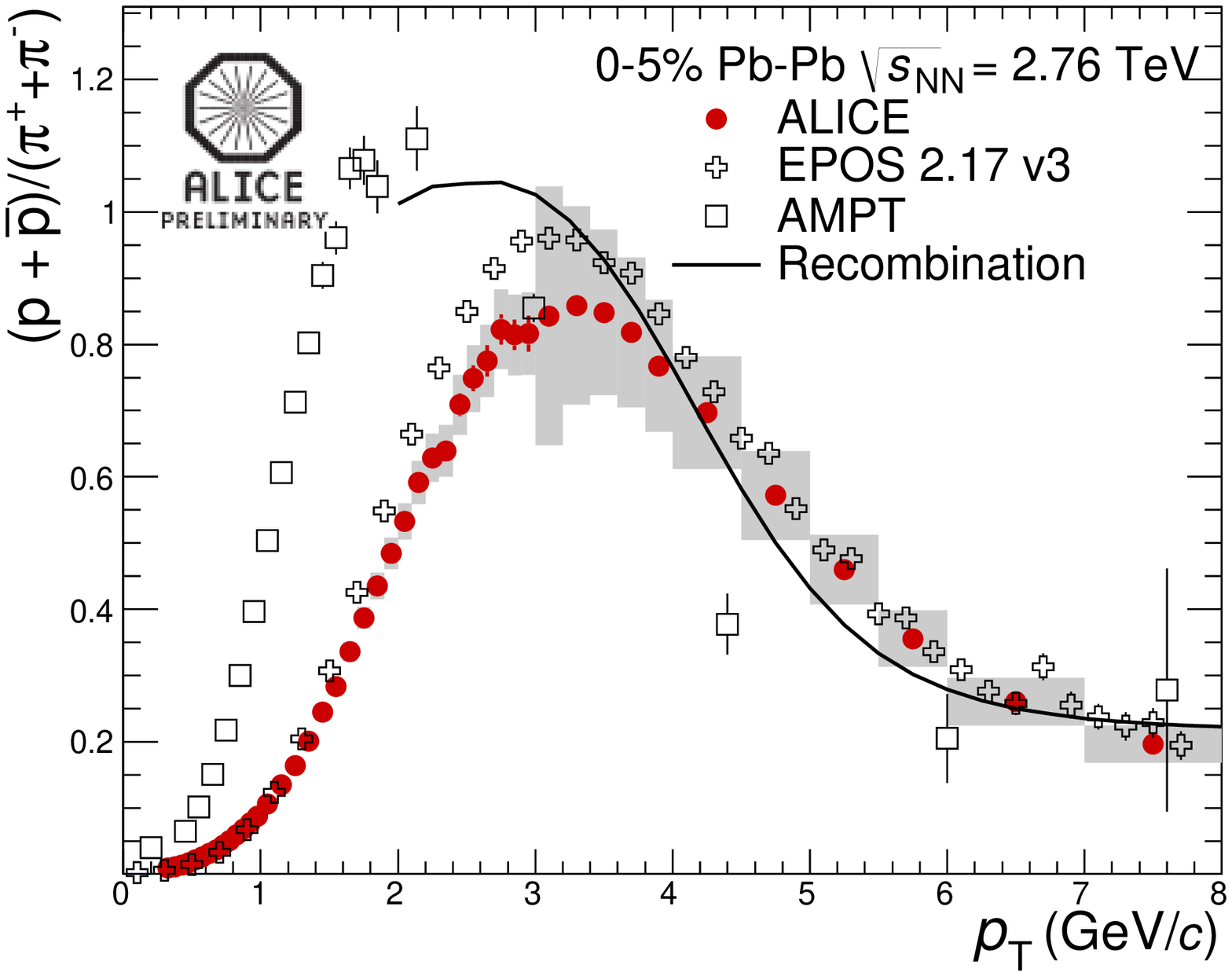}
  \includegraphics[width=.305\linewidth]{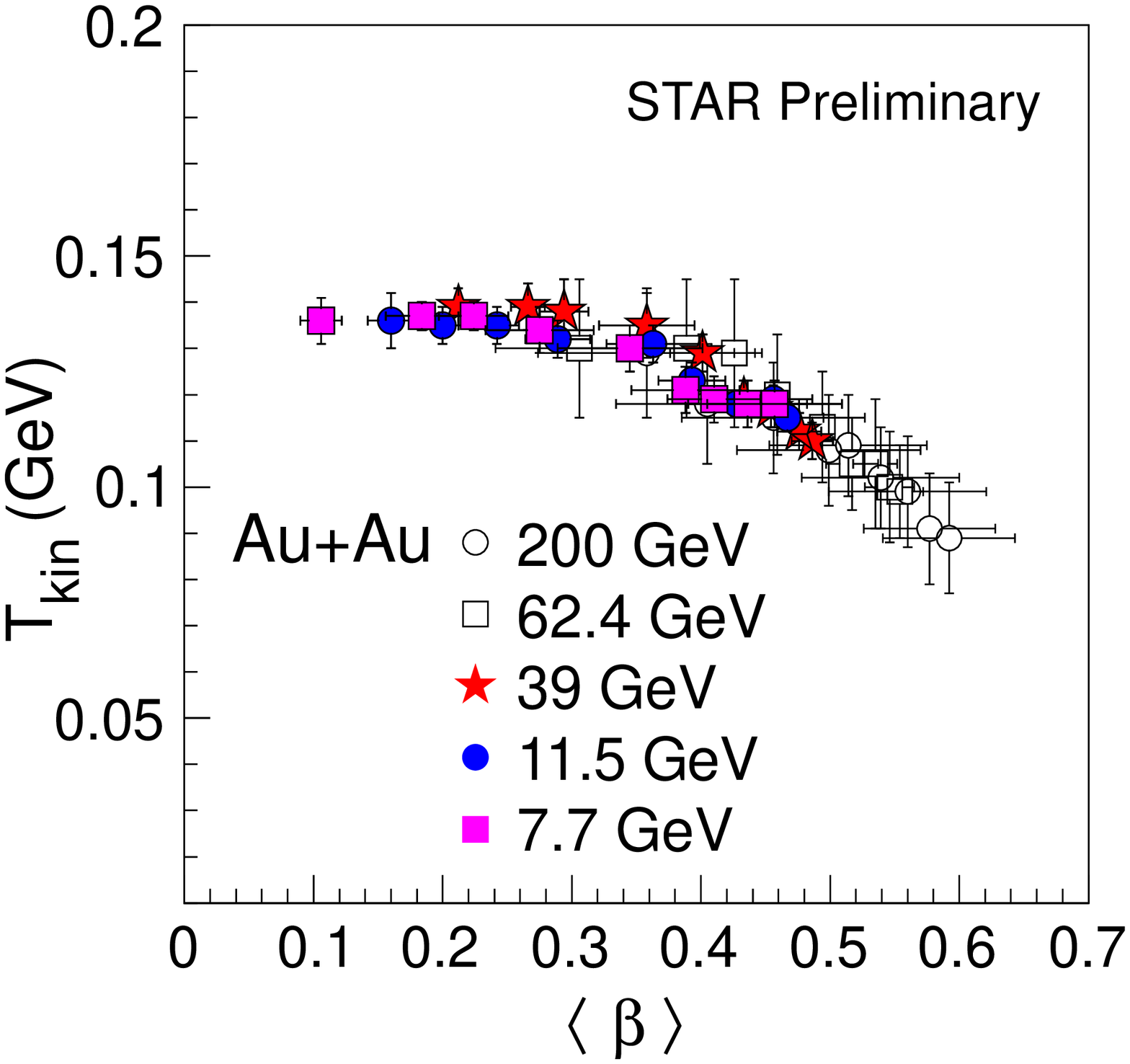}
  \end{center}
  \caption{Left: comparison of hadron integrated yields to pion ratios at mid-rapidity for RHIC and LHC central
  collisions with thermal model predictions~\cite{Milano:2012qm}.
  Center: comparison between the $\mathrm{p}/\pi$  ratio as a function of $\pT$ for 0--5$\%$ central Pb--Pb
  collisions at $\sqrtSnn=2.76~\tev$ and several models~\cite{OrtizVelasquez:2012te}.
  Right: Evolution of $\Tkin$ vs.\ $\meanBetaT$ extracted with simultaneous blast-wave fits to $\pi^{\pm}$,
  $\mathrm{K}^{\pm}$, $\mathrm{p}$, and $\mathrm{\overline{p}}$ $\pT$-spectra for different centrality intervals
  and beam energies at RHIC~\cite{Das:2012qm}.}
  \label{fig:chemical_coal_thermal}
  \end{figure}

  Once the QGP cools down and chemical freeze-out occurs, hadron abundances
  are very close to the ones recorded with the experimental setups.
  Statistical thermal model analyses of the hadron yields have been successfully describing
  the measurements on a large range of beam energies with few global variables only: the
  chemical freeze-out temperature $\Tch$, the baryo-chemical potential $\muB$ and the
  volume $V$ of the fireball.
  The STAR analyses of BES data at RHIC offer the possibility to investigate not only
  the beam-energy but also the centrality dependence of these parameters~\cite{Das:2012qm}.
  For the lowest energies of $\sqrtSnn=11.5$ and $7.7~\gev$, the results of the fits for peripheral
  collisions seem to systematically depart\footnote{Both grand canonical and strangeness canonical
  ensembles are used within the THERMUS~\cite{Wheaton:2004qb} code, which lead to opposite
  variations of the extracted $\Tch$ parameter.}
  from the phenomenological description of $1~\gev$ per hadron~\cite{Cleymans:1998fq}.
  The parameter extrapolation from RHIC to LHC energies is straightforward, since one expects $\Tch$
  to get closer to the deconfinement temperature $\Tc$ of lQCD and $\muB$ to vanish with increasing
  beam energy.
  However, some tension appears with the ALICE data analysis~\cite{Milano:2012qm}: proton and
  anti-proton yields are low with respect to calculations for a temperature of $\Tch=164~\mev$ (see
  Fig.~\ref{fig:chemical_coal_thermal}, left panel) which may reflect the presence of annihilation during
  the hadronic phase.
  The importance of corrections for feed-down and secondaries from interactions with the detector material
  is underlined together with the benefit of vertex detectors for this purpose~\cite{Andronic:2012dm}.
  Hypertriton and anti-hypertriton are measured not only at RHIC but also at the LHC.
  Using the BES data as well as the additional statistics of the 2012 run, the STAR Collaboration manages
  to obtain the excitation function of the $^3_{\Lambda}\mathrm{H}$ +$^3_{\Lambda}\overline{\mathrm{H}}$
  $\pT$-spectra and refines the estimate of the lifetime $\tau$ with measurements in a larger decay length interval.
  A combined fit of 2010+2012 results gives: $\tau=138\pm^{23}_{60}~\mathrm{ps}$~\cite{Zhu:2012qm}.
  Although ALICE extracts clear signals for $^3_{\Lambda}\mathrm{H}$, $^3_{\Lambda}\overline{\mathrm{H}}$
  and $^4_{\Lambda}\overline{\mathrm{He}}$, no H-dibaryon peak is observed and therefore, upper limits on
  the yields of this hypothetical particle are derived~\cite{Doenigus:2012qm}.
  
  The observation of baryon/meson ratios at intermediate $\pT$ being significantly higher for top RHIC energy
  Au--Au collisions as compared to pp triggered the development of models recombining quarks from the QGP
  phase to produce hadrons.
  STAR tried to isolate the possible onset of parton recombination looking at the hyperon production using the
  BES program and in particular the $\Omega/\phi$ ratio vs.\ $\pT$: again, a difference seems to be seen for
  the lowest beam energies $\sqrtSnn~=~11.5~\gev$~\cite{Zhang:2012qm}.
  As illustrated by the $\mathrm{p}/\pi$ $\pT$-ratio as a function of the Pb--Pb collision centrality in
  Fig.~\ref{fig:chemical_coal_thermal} (center panel), the enhancement observed at RHIC still holds at LHC
  energies and the most central value (0--5$\%$) is qualitatively compatible with several models using
  hadronisation mechanisms different from mere fragmentation~\cite{OrtizVelasquez:2012te}.
  A mapping of the kinetic freeze-out temperature $\Tkin$ vs.\ the mean transverse velocity $\meanBetaT$
  parameters is obtained performing blast-wave fits on $\pi^{\pm}$, $\mathrm{K}^{\pm}$, $\mathrm{p}$,
  and $\mathrm{\overline{p}}$ $\pT$-spectra~\cite{Das:2012qm,Milano:2012qm}.
  Both beam energy and centrality dependences are investigated with collisions from
  $\sqrtSnn~=~7.7~\gev$ to $2.76~\tev$.
  The evolution of the estimated radial flow is smooth with $\meanBetaT$ increasing from $\sim0.1$
  to $0.65$ and simultaneously $\Tkin$ decreasing from $\sim135$ to below $100~\mev$ (see
  Fig.~\ref{fig:chemical_coal_thermal}, right panel).
  For the most central collisions at the LHC, fluid-dynamical models are in general in good agreement with
  the identified $\pT$-spectra when a hadronic phase is included, in particular with antibaryon-baryon annihilation.
  The $\pT$-spectra corresponding to $100$~events simulated with MUSIC~\cite{Ryu:2012at}+UrQMD~\cite{Bass:1998ca,Bleicher:1999xi} during the time of the conference are very close to the measurements~\cite{Milano:2012qm}.
  %%--------------------
  %%--------------------
%%-----------------------------------------------------------------------------
\section{Future prospects}
  \label{sec:future_prospects}
  In the very near future, we perceive it to be mandatory to scrutinize the new measurements and check the
  consistency of the experimental results.
  Models have to be validated in the energy range from $\sqrt{s}=7.7~\gev$ up to $2.76~\tev$.
  Then, systematic studies to further constrain the initial-state fluctuations are necessary.
  The hope is to get a better handle from detailed EbyE harmonic flow analyses, including event-plane angles.
  In particular, ultra-central events may serve as an interesting testing ground for models of the initial-state fluctuations.

  On the theory side, further developments point into the direction of including thermal fluctuations into the
  fluid-dynamical modelling, see Ref.\ \cite{flucs}.
  Of particular importance here seems to clarify how such fluctuations enter the dynamical equations of transient
  (second-order) fluid dynamics.
  Other interesting developments are the consistent description of the dynamics of the chiral order parameter
  within a fluid-dynamical framework \cite{Nahrgang:2012qm}.
  Another interesting task for the future is the development of chiral anomalous fluid dynamics in order to assess
  the chiral magnetic effect~\cite{Kharzeev:2007jp} and the charge asymmetry observed by STAR~\cite{Wang:2012qm} and
  ALICE~\cite{Hori:2012qm}.

  Finally, explaining possible changes of hadron yields after chemical freeze-out (for instance, due to
  baryon-antibaryon annihilation) and reconciling the statistical thermal model with data also seems to
  be an important task for future studies.
%%-----------------------------------------------------------------------------
% \section*{References}
  %%

\end{document}